\documentclass[12pt]{article}
\usepackage{amsfonts}

\begin{document}

\title{Deformed Heisenberg algebra with minimal length and equivalence principle}

\author{V. M. Tkachuk\\
Department for Theoretical Physics,\\ Ivan Franko National
University of Lviv,\\ 12 Drahomanov St., Lviv, UA-79005, Ukraine\\
e-mail: tkachuk@ktf.franko.lviv.ua\\
voltkachuk@gmail.com}

\maketitle

\begin{abstract}
Studies in string theory and quantum gravity lead to the
Generalized Uncertainty Principle (GUP) and suggest the existence of
a fundamental minimal length which, as was established,
can be obtained within the deformed Heisenberg algebra.
The first look on the classical motion of bodies in a space with corresponding deformed Poisson brackets
in a uniform gravitational field can give an impression that bodies of different
mass fall in different ways and thus the
equivalence principle is violated.
Analyzing  the kinetic energy of a composite body we find that the motion of its center of mass in the deformed space
depends on some effective parameter of deformation. It gives a possibility to recover
the equivalence principle in the space with deformed Poisson brackets
and thus GUP is reconciled with the equivalence principle.
We also show that the independence of kinetic energy on composition leads to the recovering of the equivalence principle in the space with deformed Poisson brackets.
\end{abstract}

\section{Introduction}
Recently, lots of attention has been devoted to studies of different systems in a space with
a deformed Heisenberg algebra that takes into account the quantum nature of space on the phenomenological level.
These works are motivated
by several independent lines of investigations in string theory and quantum gravity (see, e.g., \cite{gross, maggiore, witten}) which lead to the
Generalized Uncertainty Principle (GUP)
\begin{eqnarray}
\Delta X\ge{\hbar\over2}\left({1\over \Delta P}+\beta\Delta P\right)
\end{eqnarray}
and suggest the existence of the
fundamental minimal length $\Delta X_{\rm min}=\hbar\sqrt\beta$, which is
of order of Planck's length $l_p=\sqrt{\hbar G/c^3}\simeq 1.6\times 10^{-35}\rm m$.

It was established that minimal length
can be obtained in the frame of small quadratic modification (deformation) of the Heisenberg algebra \cite{Kem95,Kem96}
\begin{eqnarray}
[X,P]=i\hbar(1+\beta P^2).
\end{eqnarray}
In the classical limit $\hbar\to 0$ the quantum-mechanical commutator for operators is replaced by the Poisson bracket for corresponding classical variables
\begin{eqnarray}
{1\over i\hbar}[X,P]\to\{X,P\},
\end{eqnarray}
which in the deformed case reads
\begin{eqnarray}
\{X,P\}=(1+\beta P^2).
\end{eqnarray}

 We point out that historically the first algebra of that kind in the relativistic case was proposed by Snyder in 1947 \cite{Snyder47}. But only after investigations in string theory and quantum gravity the considerable interest in the studies of physical properties of classical and quantum systems in spaces with deformed algebras appeared.

Observation that GUP can be obtained from the deformed Heisenberg algebra opens the possibility to study the influence of minimal length on properties of  physical systems on the quantum level as well as on the classical one.

Deformed commutation relations bring new difficulties in the quantum
mechanics as well as in the classical one. Only a few problems are known to be solved exactly.
They are: one-dimensional harmonic
oscillator with minimal uncertainty in position \cite{Kem95} and
also with minimal uncertainty in position and momentum
\cite{Tkachuk1,Tkachuk2}, $D$-dimensional isotropic harmonic
oscillator \cite{chang, Dadic}, three-dimensional Dirac oscillator
\cite{quesne},
(1+1)-dimensional Dirac
oscillator within Lorentz-covariant deformed algebra \cite{Quesne10909},
one-dimensional Coulomb problem
\cite{fityo}, and
the
singular inverse square
potential with a minimal length \cite{Bou1,Bou2}.
Three-dimensional
Coulomb problem with deformed Heisenberg algebra was studied within the perturbation theory \cite{Brau,Benczik,mykola,Stet,mykolaOrb}.
In \cite{Stet07} the scattering problem in the deformed space with minimal length was studied.
The ultra-cold
neutrons in gravitational field with minimal length were considered in
\cite{Bra06,Noz10,Ped11}.
The influence of minimal length on Lamb's shift, Landau levels, and tunneling current in scanning tunneling   microscope was studied \cite{Das,Ali2011}.
The Casimir effect in a space with minimal length was examined in \cite{Frassino}.
In \cite{Vaki} the effect of noncommutativity and of the existence of a minimal length on the phase space of cosmological model was investigated.
The authors of paper \cite{Batt}
studied various physical consequences which follow from the noncommutative Snyder space-time geometry.
The classical mechanics in a space with deformed Poisson brackets was studied
in \cite{BenczikCl,Fryd,Sil09}.
The composite system ($N$-particle system) in the deformed space with
minimal length was studied in \cite{Quesne10,Bui10}.

Note that deformation of Heisenberg algebra brings not only technical difficulties in solving of corresponding equations
but also brings problems of fundamental nature.
One of them is the violation of the equivalence principle in
space with minimal length \cite{Ali11}.
This is the result of assumption that the parameter of deformation
for
macroscopic bodies of different mass is unique.
In paper \cite{Quesne10} we shown that the center of mass of a macroscopic body in deformed space is
described by an effective parameter of deformation, which is essentially smaller than the parameters of deformation for particles consisting the body. Using the result of \cite{Quesne10} for the effective parameter of deformation we show that the equivalence principle in the space with minimal length can be recovered.
In section 3 we reproduce the result of \cite{Quesne10} concerning the effective parameter of deformation for the center of mass on the classical level and in addition show that the independence of kinetic energy on the composition leads to the recovering of the equivalence principle in the space with deformed Poisson bracket.

\section{Free fall of particle in a uniform gravitational field}

The Hamiltonian of a particle (a macroscopic body which we consider as a point particle) of mass $m$ in a uniform gravitational field reads
\begin{eqnarray}
H={P^2\over 2m}-mgX,
\end{eqnarray}
the gravitational field is characterized by the factor $g$ is directed along the $x$ axis.
Note that here the inertial mass ($m$  in the first term) is equal to the gravitational mass
($m$ in the second one).
The Hamiltonian equations of motion in space with deformed Poisson brackets are as follows
\begin{eqnarray}\label{dxp}
\dot{X}=\{X,H\}={P\over m}(1+\beta P^2),\\
\dot{P}=\{P,H\}=mg(1+\beta P^2).
\end{eqnarray}
We impose zero initial conditions for position and momentum, namely $X=0$, and $P=0$ at $t=0$.
These equations can be solved easily.
From the second equation we find
\begin{eqnarray}
P={1\over \sqrt\beta}\tan(\sqrt\beta mgt).
\end{eqnarray}
From the first equation  we obtain for velocity
\begin{eqnarray}\label{soldX}
\dot{X}={1\over m\sqrt\beta}{\tan(\sqrt\beta mgt)\over\cos^2(\sqrt\beta mgt)}
\end{eqnarray}
and for position
\begin{eqnarray}\label{solX}
X={1\over 2g m^2\beta}\tan^2(\sqrt\beta mgt).
\end{eqnarray}
One can verify that the motion is periodic with period $T={\pi\over m\sqrt\beta g}$. The particle moves from $X=0$
to $X=\infty$, then reflects from $\infty$ and moves in the opposite direction to $X=0$.
But from the physical point of view this solution is correct only for time $t\ll T$ when the velocity of particle
is much smaller than the speed of light. In other cases, the relativistic mechanics must be considered.

It is instructive to write out the results for velocity and coordinate in the first order over $\beta$:
\begin{eqnarray}\label{soldXap}
\dot{X}=gt\left(1+{4\over 3}\beta m^2g^2t^2\right),\\\label{solXap}
X= {gt^2\over2}\left(1+{2\over 3}\beta m^2g^2t^2\right).
\end{eqnarray}
In the limit
$\beta\to 0$ we reproduce the well known results
\begin{eqnarray}
\dot{X}=gt, \ \
X= {gt^2\over2},
\end{eqnarray}
where kinematic characteristics, such as velocity and position of a free-falling particle depend only on initial position and velocity of the particle and do not depend on the composition and mass of the particle.
It is in agreement with the weak equivalence principle, also known as the universality of free fall or the Galilean equivalence principle. Note that in the nondeformed case, when the Newtonian equation of motion in gravitational field is fulfilled the weak equivalence principle is noting else that the statement of equivalence
of inertial and gravitational masses.

As we see from (\ref{soldX}) and (\ref{solX}) or (\ref{soldXap}) and (\ref{solXap}), in the deformed space
the trajectory of the point mass in the gravitational field depends on the mass of the particle if we suppose that
parameter of deformation is the same for all bodies.
So, in this case the equivalence principle is violated.
In paper \cite{Quesne10} we shown on the quantum level that in fact the motion of the center of mass of a composite system in deformed space is governed by an
effective parameter (in \cite{Quesne10} it is denoted as $\tilde\beta_0$, here we denote it as $\beta$). So, the parameter of deformation for a macroscopic body
is
\begin{eqnarray}\label{betaN}
\beta=\sum_i\mu_i^3\beta_i,
\end{eqnarray}
where
$\mu_i=m_i/\sum_i m_i$, $m_i$ and $\beta_i$ are the masses and parameters of deformation of particles which form composite system (body). Note that in the next section we derive this result considering kinetic energy of a body consisting of $N$ particles.

Firstly, let us consider a special case $m_i=m_1$ and $\beta_i=\beta_1$ when body consists of the same elementary particles. Then we find
\begin{eqnarray}
\beta={\beta_1\over N^2},
\end{eqnarray}
where $N$ is the number of particles of body with mass $m=Nm_1$.
Note that expressions (\ref{soldX}) and (\ref{solX}) contain combination $\sqrt\beta m$.
Substituting the effective parameter of deformation
$\beta_1/N^2$ instead of $\beta$ we find
\begin{eqnarray}
\sqrt\beta m=\sqrt\beta_1 m/N=\sqrt\beta_1 m_1.
\end{eqnarray}
As a result, the trajectory now does not depend on the mass of the macroscopic body but depends on
$\sqrt\beta_1 m_1$, which is the same for bodies of different mass.
So, the equivalence principle is recovered.

The general case when a body consists of the different elementary particles is more complicated.
Then the situation is possible when different combinations of elementary particles
lead to the same mass but with different effective parameters of deformation.
Then the motion of bodies of equal
mass but different composition will be different.
This also violates the weak equivalence principle.
The equivalence principle can be recovered when we suppose that
\begin{eqnarray}\label{gamma}
\sqrt\beta_1 m_1=\sqrt\beta_2 m_2=\dots=\sqrt\beta_N m_N=\gamma
\end{eqnarray}
Really, then the effective parameter of deformation for a macroscopic body is
\begin{eqnarray}
\beta=\sum_i{m_i^3\over(\sum_i m_i)^3}\beta_i={\gamma^2\over(\sum_i m_i)^2}={\gamma^2\over m^2}
\end{eqnarray}
and thus
\begin{eqnarray}
\sqrt\beta m=\gamma,
\end{eqnarray}
that is the same as (\ref{gamma}).
Note, that the trajectory of motion in this case does not depend on mass and depends only on $\gamma$
which takes same value for all bodies.
It means that bodies of different mass and different composition move in a gravitational field in the same way
and thus the weak equivalence principle is not violated when (\ref{gamma}) is satisfied. Equation  (\ref{gamma}) brings one new fundamental constant $\gamma$. Note that parameter $1/\gamma$ has the dimension of velocity.  The parameters of deformation $\beta_i$ of particles or macroscopic bodies of mass $m_i$ are determined by fundamental constant $\gamma$ as follows
\begin{eqnarray}\label{bg}
\beta_i={\gamma^2\over m_i^2},
\end{eqnarray}
So, the parameter of deformation is completely determined by the mass of a particle.
In the next section we derive formula (\ref{betaN}) on the classical level and give some arguments concerning
the relation (\ref{gamma}).

\section{Kinetic energy of a composite system in deformed space and parameter of deformation}

In this section we use the natural statement:
{\it The kinetic energy has the additivity property and does not depend on composition of a body but only on its mass.}

Firstly, we consider {\it the additivity property of the kinetic energy.}
Let us consider $N$ particles with masses $m_i$ and deformation parameters $\beta_i$.
It is equivalent to the situation when the macroscopic body is divided into $N$ parts which can be treated as point particles with corresponding masses and parameters of deformation.
We consider the case when each particle of the system moves with the same velocity as the whole system.

Let us rewrite the kinetic energy as a function of velocity.
From the relation between velocity and momentum (\ref{dxp}) in the first approximation over $\beta$
we find
\begin{eqnarray}
P=m \dot X(1-\beta m^2\dot X^2).
\end{eqnarray}
Then the kinetic energy as a function of velocity in the first order approximation over $\beta$ reads
\begin{eqnarray}\label{TV}
T={m\dot X^2\over 2}-\beta m^3\dot X^4.
\end{eqnarray}

The kinetic energy of the whole system is given by (\ref{TV}) where $m=\sum_i m_i$. On the other hand,
the kinetic energy of the whole system is the sum of kinetic energies of particles which constitute the system:
\begin{eqnarray}\label{TVsum}
T=\sum_i T_i={m\dot X^2\over 2}-\sum_i\beta_i m_i^3\dot X^4,
\end{eqnarray}
where we take into account that velocities of all particles are the same as the velocity
of the whole system $\dot X_i=\dot X$, $i=1,\dots,N$.
Comparing (\ref{TV}) and (\ref{TVsum}) we obtain (\ref{betaN}).

Now let us consider {\it the independence of kinetic energy on the composition of a body}.
It is enough to consider a body of a fixed mass consisting of two parts (particles) with masses $m_1=m\mu$ and $m_2=m(1-\mu)$, where $0\le\mu\le1$. Parameters of deformation for the first and second particles are $\beta_1=\beta_{\mu}$ and $\beta_2=\beta_{1-\mu}$, here we write explicitly that
parameters of deformations are some function of mass ($\mu=m_1/m$ is dimensionless mass).
The particles with different masses constitute the body with the same mass $m=m_1+m_2$.
So, in this situation we have the body of the same mass but with different composition.

The kinetic energy of the whole body is given by (\ref{TV}) with the
parameter of deformation
\begin{eqnarray}\label{Eqbeta}
\beta=\beta_{\mu}\mu^3+\beta_{1-\mu}(1-\mu)^3.
\end{eqnarray}
Since the kinetic energy does not depend on the composition, the parameter of deformation for the whole body must be fixed $\beta={\rm const}$ for different $\mu$. Thus (\ref{Eqbeta}) is the equation for $\beta_{\mu}$ as a function of $\mu$ at fixed $\beta$.
One can verify that the solution reads
\begin{eqnarray}
\beta_{\mu}={\beta\over\mu^2}.
\end{eqnarray}
Taking into account that $\mu=m_1/m$ we find
\begin{eqnarray}
\beta_1 m_1^2=\beta m^2
\end{eqnarray}
that corresponds to (\ref{gamma}). So, the independence of the kinetic energy from composition leads to the one fundamental constant $\gamma^2=\beta m^2$. Then parameters of deformation $\beta_i$ of particles or composite bodies
of different masses $m_i$ are
$\beta_i=\gamma^2/m_i^2$
that is in agreement with relation (\ref{bg}).
\section{Conclusions}
One of the main results of the paper is the expression for the parameter of deformation
for particles or bodies of different mass (\ref{bg})
which recovers the equivalence principle and thus the equivalence principle is reconciled with the
generalized uncertainty principle. It is necessary to stress that
expression (\ref{bg}) was derived also in section 3 from the
condition of the independence of kinetic energy on composition.

Note that (\ref{bg}) contains the same constant $\gamma$ for different particles and parameter of deformation
is inverse to the squared mass.
The constant $\gamma$ has dimension inverse to velocity. Therefore, it is convenient to introduce
a dimensionless constant $\gamma c$, where $c$ is the speed of light.
In order to make some speculations concerning the possible value of $\gamma c$
we suppose that for the electron the parameter of deformation $\beta_e$ is related to Planck's
length, namely
\begin{eqnarray}
\hbar\sqrt\beta_e=l_p=\sqrt{\hbar G/c^3}.
\end{eqnarray}
Then we obtain
\begin{eqnarray}
\gamma c=c\sqrt\beta m_e=\sqrt{\alpha{Gm_e^2\over e^2}}\simeq 4.2\times 10^{-23},
\end{eqnarray}
where $\alpha=e^2/\hbar c$ is the fine structure constant.

Fixing the parameter of deformation for electron we can calculate the
parameter of deformation for particles or bodies of different mass. It is more instructive to write
the minimal length for space where the composite body of mass $m$ lives:
\begin{eqnarray}
\hbar\sqrt\beta={m_e\over m}\hbar\sqrt\beta_e={m_e\over m }l_p.
\end{eqnarray}
As an example let us consider nucleons (proton or
neutron). The parameter of deformation for nucleons $\beta_{\rm nuc}$ or minimal length for nucleons
reads
$\hbar\sqrt\beta_{\rm nuc}\simeq  l_p/1840.$
So, the effective minimal length for nucleons is three order smaller than that for electrons.

\end{document}